\documentclass[conference]{IEEEtran}
\IEEEoverridecommandlockouts

\usepackage{cite}
\usepackage{amsmath,amssymb,amsfonts}
\usepackage{algorithmic}
\usepackage{graphicx}
\usepackage{xcolor}
\usepackage{multirow}
\usepackage{textcomp}
\usepackage{bm}
\usepackage[hidelinks]{hyperref}
\usepackage[capitalize]{cleveref}
\usepackage{relsize}
\usepackage{booktabs}
\renewcommand{\baselinestretch}{0.985}
\usepackage{color_schemes}
\def\BibTeX{{\rm B\kern-.05em{\sc i\kern-.025em b}\kern-.08em
    T\kern-.1667em\lower.7ex\hbox{E}\kern-.125emX}}
\begin{document}

\title{Physics-Informed Neural Networks in Power System Dynamics: Improving Simulation Accuracy\\

\thanks{I. Ventura Nadal, R. Nellikkath and S. Chatzivasileiadis were supported by the ERC Starting Grant VeriPhIED, funded by the European Research Council, Grant Agreement 949899.}
}

\author{\IEEEauthorblockN{Ignasi Ventura Nadal, Rahul Nellikkath, and Spyros Chatzivasileiadis}
\IEEEauthorblockA{Department of Wind and Energy Systems \\
\textit{Technical University of Denmark}\\
Kongens Lyngby, Denmark \\
\{ignad, rnelli, spchatz\}@dtu.dk}
}


\maketitle

\begin{abstract}
The importance and cost of time-domain simulations when studying power systems have exponentially increased in the last decades. With the growing share of renewable energy sources, the slow and predictable responses from large turbines are replaced by the fast and unpredictable dynamics from power electronics. The current existing simulation tools require new solutions designed for faster dynamics. Physics-Informed Neural Networks (PINNs) have recently emerged in power systems to accelerate such simulations. By incorporating knowledge during the up-front training, PINNs provide more accurate results over larger time steps than traditional numerical methods. This paper introduces PINNs as an alternative approximation method that seamlessly integrates with the current simulation framework. We replace a synchronous machine for a trained PINN in the IEEE 9-, 14-, and 30-bus systems and simulate several network disturbances. Including PINNs systematically boosts the simulations' accuracy, providing more accurate results for both the PINN-modeled component and the whole multi-machine system states.

\end{abstract}

\section{Introduction}
Time-domain simulations are routinely used in power systems to ensure a stable and reliable operation. These simulations provide critical insights to study and analyse the system's response to a wide range of operating conditions and disturbances \cite{kundur, sauerandpi}. The current simulation tools are facing major and growing computational challenges. The green transition is shifting the power systems operating paradigm, resulting in faster and more distributed grid dynamics. The established simulating frameworks, although they are modular and numerically stable, rapidly become very computationally expensive when studying fast dynamics \cite{stott}.

Thus, over the past few decades, a big effort in the power system community has been centred on accelerating simulation platforms. Well-known examples of such effort are more computing resources, more optimized algorithms \cite{aristidou}, and model order reduction methodologies \cite{modelreduction}. Although these efforts have improved the efficiency of the simulation platforms, they are still constrained by the fundamental theory. In order to achieve significant computation advantage for time-domain simulations, we need to rethink core modeling decisions.

In recent years, Neural Networks (NNs) have revolutionized multiple fields. By learning from data, they are capable of approximating any given continuous function \cite{universal}. The so-called Physics-Informed Neural Networks (PINNs), which incorporate the governing physics during training, were recently shown to learn any given system of differential equations (ODEs) \cite{raissi}. Yielding high-speed and sufficiently accurate results over long time steps, a trained PINN offers an alternative to the classical methods, as it is not constrained by small time steps to ensure accuracy and numerical stability \cite{resnet, piml, ann}. These characteristics make PINNs extremely attractive for power system dynamic simulations. In the power systems field, they were first used by \cite{misyris} to capture the so-called swing equation of a single-machine infinite bus system.

A new wave of power systems research followed to harness the PINN characteristics in time-domain simulations \cite{apppinns}. The first approaches focused on capturing whole system dynamics with one model, quickly demonstrating simulation speed-ups of up to four orders of magnitude \cite{jochenepsr, daepinn, kyri}. However, high up-front training costs and poor generalisation made this an unfeasible implementation strategy. The next approaches tried to overcome these scalability and generalistion issues by training for individual components. The authors in \cite{pinnsim} introduced a novel simulator where each component is captured by PINNs and later interfaced with a root-finding algorithm. To remove barriers for the adoption of PINNs in power system dynamic simulations, \cite{plugandplay} introduced a parametrization compatible with the established simulation frameworks which can enable the seamless integration of PINNs to existing tools. In that work, terminal voltages were considered as the interfacing variables between components.

This paper offers a new perspective on how PINNs can be parametrized to better capture dynamic components and be integrated into the established simulation framework, boosting its accuracy and speed performance. The formulated PINNs capture the component dynamics considering the system's interaction through the internal circuit variables, i.e. the injected currents. This formulation is compatible and non-exclusive with other integration methods, allowing simulations to model different components with both PINNs and traditional methods simultaneously. It further enables a better synergy between time-domain simulations and emerging deep learning techniques, benefiting from both frameworks' characteristics. The contributions of this paper are as follows:

\begin{itemize}
    \item We propose a Physics-Informed Neural Networks (PINNs) formulation as a more accurate alternative to the classic numerical methods used in power system simulation frameworks. Using injected currents to interface with the system, the formulation learns the dynamics of individual components.
    \item In the IEEE 9-, 14-, and 30-bus test systems, we capture the machine having the lowest inertia with a PINN and simulate the system's response to different network disturbances. We show how introducing a PINN significantly improves the overall accuracy of the simulation.
\end{itemize}

We post our code online \cite{github_PINN}, so that the power systems community can (i) use our framework to boost their dynamic simulations with PINNs, or (ii) train their PINNs for their specific components.

The remainder of this paper is structured as follows. In Sec. \ref{sec2}, we formulate the time-domain simulation problem and the current limitations. In Sec. \ref{sec3}, we propose PINNs as an accurate alternative to traditional integration methods. Test results, collected from the IEEE 9-, 14-, and 30-bus systems are presented in Sec. \ref{sec4}. Conclusions are offered in Sec. \ref{sec5}.

\section{Power System Modeling and Opportunities} \label{sec2}
In this section, we formulate the time-domain simulation problem, describe the numerical integration techniques used for the differential equations, and define the algorithm used to carry out the simulation iteratively. We consider the phasor mode approximation, describing how the phasors evolve with time and neglecting electromagnetic phenomena.

\subsection{Time-domain simulations problem}
An electric power system consists of a set of dynamic components, described by a system of nonlinear differential equations, that interact inside a shared network, defined by the power flow equations. This system of Differential and Algebraic Equations (DAE) has the form
\begin{subequations}\label{daesystem}
\begin{align}
        \frac{d}{dt}x &= f(x, \Bar{I}, \Bar{V}, u) \label{daedif} \\
        \Bar{I} &= h(x, \Bar{V}) \label{daecurrents}\\
        0 &= g(x, \Bar{I}, \Bar{V}), \label{daenetwork}
\end{align}
\end{subequations}
with the component state variables $x(t) \in \Re^n$, the complex bus voltages $\Bar{V}(t) \in \Re^m$, the complex injected currents $\Bar{I}(t) \in \Re^n$, and inputs $u(t) \in \Re^n$. The states update function $f$: $\Re^{n+m} \Rightarrow \Re^n$, and the component's internal circuits and network equations $h$: $\Re^{n+m} \Rightarrow \Re^n$ and $g$: $\Re^{n+m} \Rightarrow \Re^m$, respectively. We compute the initial variable values by solving the power flow equations and then solve the system over time.

\subsection{Numerical integration of the differential equations}

The system described by \eqref{daesystem} cannot be solved analytically; thus, we need to integrate the differential equations to obtain the explicit values of $x$. This integration is described in \eqref{eq:state_integration}, where we go from the initial value $x_n$ to a future value $x_{n+\Delta t}$:
\begin{equation}\label{eq:state_integration}
    x_{n+\Delta t} = x_n + \int^{t_n + \Delta t}_{t_n} f(x, \Bar{I}, \Bar{V}, u) dt.
\end{equation}
Since we cannot compute the exact integration, we rely on numerical methods to approximate it. The approximation can be expressed with an explicit or implicit algebraic relationship, which, applied to \eqref{daesystem}, yields a system of nonlinear algebraic equations that can be solved. 

We assume the system is characterized as a semi-explicit DAE of index-1, which is assured for most practical cases \cite{index1}. This allows the discretization of the differential equations with classical numerical methods.

The available integration methods fall into two categories: \textit{explicit} and \textit{implicit}. In this work, we consider implicit methods from the Runge-Kutta (RK) family, which assume a polynomial evolution of the state variables $x$ in \eqref{eq:state_integration}. Among these methods, the trapezoidal rule is the most widely implemented, and extensively used in most commercial simulation software. Assuming the update functions to only depend on states and injected currents, the resulting system becomes
\begin{subequations}\label{daesystemtrapz}
\begin{align}
        x_{n+\Delta t} &= x_n + \frac{h}{2} (f(x_n,\Bar{I}_{n}) + f(x_{n+\Delta t}, \Bar{I}_{n+\Delta t})) \label{daetrapzdif} \\
        \Bar{I}_{n+\Delta t} &= h(x_{n+\Delta t}, \Bar{V}_{n+\Delta t}) \\
        0 &= g(x_{n+\Delta t}, \Bar{V}_{n+\Delta t}),
\end{align}
\end{subequations}
which can now be analytically solved as a standard nonlinear system of equations. The simulation is then solved iteratively with the specified time steps, gradually building up to the desired simulation time or convergence. 

\subsection{Time step selection: accuracy v. speed}
Estimating \eqref{daedif} with \eqref{daetrapzdif} induces an approximation error. The only way to control this error is by controlling the time step size. The smaller the time step, the smaller the approximation error $\mathlarger{\varepsilon}$ incurred, tending to zero following
\begin{equation}
    \lim_{\Delta t\rightarrow0} \mathlarger{\varepsilon} (\Delta t) = 0.
\end{equation}
Reducing the time step drastically increases the simulation's accuracy. However, reducing the step size comes with a consequential computational increase, as the number of iterations needed to solve the simulation is proportional to the time step reduction. This trade-off between accuracy and simulation speed dominates all practical applications \cite{iserles}.

Until recently, power system dynamics have been simulated with fairly large step sizes. The time responses given by synchronous machines are slow and damped, as they rely on turbines with large inertia constants. This is not the case with inverter-dominated systems, whose dynamics become faster and more unpredictable. Thus, new tools are needed to enable the safe and reliable integration of new technologies.

\subsection{Opportunities for new computing methods}
Understanding the current simulation framework provides information on the limitations of the theory and the opportunities for improvement. The current simulating framework is reliable and flexible but it is computationally expensive for small time step sizes. This opens the door to new methods, with their potential characteristics listed below.
\begin{enumerate}
    \item As discussed previously, current integration techniques are limited to tiny time steps to maintain accuracy. Thus, new methods that more accurately approximate the integration can provide larger time steps without compromising accuracy. \label{first}
    \item A key characteristic relates to the modularity of the simulating framework. The dynamics of each component interact with the network and not directly with other components. Thus, the system's Jacobian presents a block structure that allows component dynamics, shown in \eqref{daedif}, to be approximated by different methods. \label{second}
\end{enumerate}
PINNs offer accurate and numerical stable differential function approximations over large time step sizes. These characteristics make it an attractive alternative to item \ref{first}. Item \ref{second} enables a modular integration of PINNs, approximating only the specified differential equations.

The following section explains the PINN formulation and how it can be seamlessly integrated into the existing simulating frameworks to boost their performance.

\section{Neural Networks Integration} \label{sec3}
This section formulates PINNs and establishes them as an alternative integration method, describing their input domain and training routine.

\subsection{General formulation}
We use a fully connected feed-forward neural network with $K$ hidden layers and $N^k$ neurons in each layer\cite{Goodfellow}. The hidden layers consist of the weight matrices $W^k$ and bias vectors $b^k$. These form the learnable parameters of the PINN, $\Theta = \{W^k, b^k\}$. All layers consider a non-linear activation function $\sigma$. Thus, the mapping from inputs to outputs is defined by 
\begin{equation}\label{eq:hiddenlayer}
    z_{k+1} = \sigma (W^{k+1} z_k + b^{k+1}), \; \forall k = 0, 1, ..., K-1.
\end{equation}

To achieve a structure similar to \eqref{eq:state_integration}, we adjust the output layer to enforce the initial conditions $x_n$ as hard constraints, and multiply the PINNs output by the step size. This approach improves numerical consistency, and effectively provides an alternative to traditional integration schemes, such as Runge-Kutta methods. Defined in \eqref{eq:outputlayer}, it provides a direct mapping between initial and final differential states,
\begin{equation} \label{eq:outputlayer}
    \hat{x}_{n+1} = x_n + \Delta t \times \big( \sigma(W^K z_K +b^K) \big).
\end{equation}

\subsection{PINN formulation}
The input domain considers all the simulation variables to capture the correct trajectories. These fall into three groups: (i) time step size $\Delta t$, (ii) initial states $x_n(t)$, and (iii) algebraic variables evolution $\Bar{I}_n(t)$, accounting for the system's interaction. \Cref{eq:outputlayer} becomes \eqref{eq:pinnformu} when considering the input domain, establishing PINNs as a method to approximate \eqref{eq:state_integration}. This formulation uses the same variables as the trapezoidal rule, previously described in \eqref{daetrapzdif}.
\begin{equation} \label{eq:pinnformu}
    \hat{x}_{n+1} = x_n + \Delta t \times\, \mathrm{PINN}(\Delta t, x_n, \Bar{I}_n, \Bar{I}_{n+h}),
\end{equation}

\begin{figure}[!ht]
    \centering
    \includegraphics[width=0.999\linewidth]{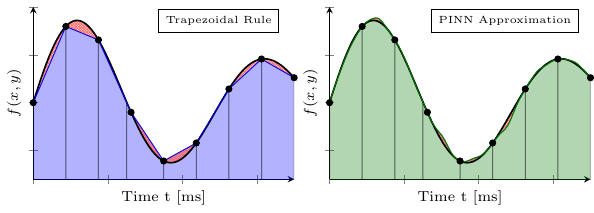}
    \caption{Depicted are the trapezoidal rule (blue) and PINN approximation (green) algorithms to estimate the integral of the differential equation \eqref{daedif}. These are compared to the exact trajectory (black line) using the same time steps (black dots). The red areas depict the errors between exact and approximated trajectories.}
    \label{fig:explanation}
\end{figure}

Unlike traditional integration schemes, the presented PINN formulation can approximate any nonlinear continuous function with a very fast evaluation speed. \Cref{fig:explanation} conceptually illustrates its performance compared to the previously introduced trapezoidal rule. The black line represents the exact dynamics of $f(x, y)$ that we are trying to approximate. The trapezoidal rule would evaluate these functions at the end of each time step, and linearly interpolate the state evolution in between. Instead, the PINN trajectory directly targets the exact trajectory. The red areas depict the difference between the exact trajectory and the approximations. The smaller the red areas, the more accurate and numerically stable the simulation is. Thus, at the cost of the PINN training time, we achieve an approximation that is accurate, fast, and whose accuracy does not rely on the step size, but on how long it was trained for. 

\subsection{Input domain}

To allow for repeated use of the same PINN throughout simulations, we need to define an input domain for the states $\mathcal{X}_x$ and $\mathcal{X}_y$ that captures the component dynamics for any given input value. This means that the operating ranges of $x_n(t)$ and $\Bar{I}_n(t)$ have to be parametrized in the PINN. For example, internal voltage inputs would be within $E'_{d-q} \in [0.94, 1.06]$, or the rotor angle within $\delta \in [-\pi, \pi]$. Thus, with a time step size defined within $\Delta t \in [0, \Delta t_{max}]$, the trained model effectively captures the evolution of the differential equation \eqref{daedif} for all possible variable values. 

The injected currents $\Bar{I}_n(t)$, which account for the network interaction, are parametrized at the beginning and end of the time step, $\Bar{I}_{n}$ and $\Bar{I}_{n+\Delta t}$. We assume a linear profile within each time step. With this input formulation, the PINN can represent the component in any simulation and operating conditions, i.e. it becomes case-independent.

\subsection{Integration}
The presented approach is fully compatibile with the established simulation frameworks. The method is a non-exclusive alternative to existing numerical schemes. It uses deep learning to improve one of the primary simulation shortcomings, which is the trade-off between accuracy and speed.

\begin{figure}[!ht]
    \centering
    \includegraphics[width=1\linewidth]{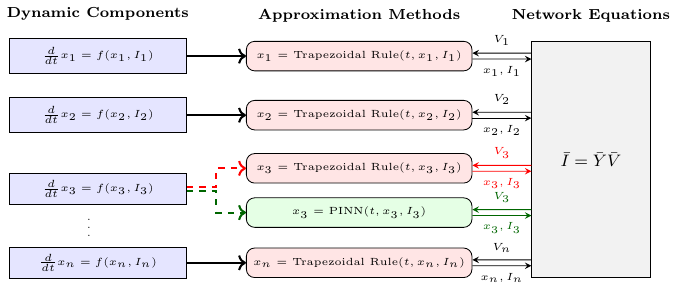}
    \caption{Represented is the power system dynamic simulations problem. Several dynamic components that share a network evolve and interact simultaneously. To solve this system, we algebraize the component dynamics with a RK method, such as the trapezoidal rule, or, as we propose in this paper, a trained PINN. The integration is modular: any component can be captured, and a simulation can include multiple components captured by a PINN.}
    \label{fig:dynamicschema}
\end{figure}

One of the primary strengths of the proposed approach is that the models are integrated into the simulation algorithms in a plug-and-play fashion. Thus, both PINNs and RK schemes can be used in the same simulation to approximate the differential states of different components. This modularity enables a smooth method integration and creates an efficient synergy between deep learning and traditional mathematical algorithms. \Cref{fig:dynamicschema} depicts how the approximation methods that connect the dynamic components with the shared network are applied and how they can be replaced by PINNs. For simplicity purposes, Fig. \ref{fig:dynamicschema} introduces a PINN as an approximation alternative only for the third dynamic component. However, any component can be approximated by a PINN, and simulations can incorporate multiple components modeled by PINNs.

\subsection{PINN training procedure}
To obtain the final trained models, i.e., the neural networks' weights and biases, we utilize a loss function containing two components: a data-based loss and a physics-based loss. The data-based loss, denoted as $\mathcal{L}_u$, uses the mean squared error to minimize the difference between PINN prediction and the target values provided by a dataset, $\mathcal{D}_u$. The dataset contains simulated states for various initial conditions, time step sizes ($h$), and algebraic evolution parameterizations.

\begin{equation}
    \mathcal{L}_u  = \frac{1}{N_u} \sum^{N_u}_{j=1} \left\Vert x^j_{n+1} - \Hat{x}^j_{n+1} \right\Vert_2^2, 
\end{equation}

Additionally, to leverage the underlying physics of the problem in the training, we use a physics-based loss $\mathcal{L}_p$:
\begin{equation}
    \mathcal{L}_p = \frac{1}{N_p} \sum^{N_p}_{j=1} 
    \left\Vert \frac{d}{dt} \hat{x}(t^j) - f\left(\hat{x}^j, \hat{y}^j\right) \right\Vert_2^2.
\end{equation}

Both data-, and physics-based losses, inspired by the formulation proposed in \cite{jochenepsr}, require the initial states, the algebraic parametrization, and the selected time step size to solve the differential equations.
The loss terms are added together with the hyper-parameter $\alpha$, as shown in \eqref{loss}. The loss function is minimized with a gradient-descent algorithm by updating the weights and biases of the chosen architecture.
\begin{subequations}
\begin{align}
    \min_{\{W^k, b^k\}_{1\leq k\leq K}} \quad &\mathcal{L}_x(\mathcal{D}_x) + \alpha \mathcal{L}_c (\mathcal{D}_c) \label{loss}\\
    \text{s.t.}\quad & \eqref{eq:hiddenlayer}, \eqref{eq:outputlayer}.
\end{align}
\end{subequations}

\subsection{Limitations of the PINN approximation}

In our current PINN application, we view the up-front training cost as a strength rather than a limitation, highlighting the method’s flexibility. This training up-front cost allows the deep learning method to learn the dynamics before the simulation,  resulting in more accurate simulations and significantly improving computational efficiency. PINNs essentially enable the incorporation of knowledge inside its method, effectively shifting a portion of the computational burden from the simulation phase to the offline training phase.

The primary limitation arises from assuming a linear evolution for the injected currents within the time step. The difference between the linear and exact profiles introduces prediction errors that grow with large step sizes. Thus, the method also has a maximum step size to ensure the required accuracy. If the assumed current profiles were more accurate, we could further increase the step sizes. We are working on better algebraic profiles to expand PINNs to larger step sizes.

Secondly, while the trained model provides significant application benefits, it also restricts its generalization to changing parameters. Thus, it is essential to ensure that the training process accounts for all relevant parameter and variable combinations encountered during the simulation.

\section{Numerical Study} \label{sec4}
This section uses the IEEE 9-, 14- and 30-bus test systems, described in \cite{illinois}, to show how PINNs can replace and work together with numerical integration schemes to improve the power system simulations accuracy. We present the case studies used, followed by the PINN implementation. Test results are then analyzed for the three test systems.

\subsection{Case study}
We consider a DAE model in the current-balance form, as it is typically implemented in most industry software \cite{sauerandpi}. \Cref{daesystemresults} adapts the formulation from \eqref{daesystem} to a standard power system, 
\begin{subequations}\label{daesystemresults}
\begin{align}
        \frac{d}{dt}x_i(t) &= f_i(x_i(t), \Bar{I}_i(t)) \; \forall i = 1, 2,\dots, N_c\label{daedifresults} \\
        \Bar{I} &= h_i(x_i(t), \Bar{V}_i(t)) \\
        0 &= g_j(x_i(t), \Bar{V}_j(t)), \; \forall j = 1, 2, \dots, N_j
\end{align}
\end{subequations}
where $N_c$ are the number of dynamic components and $N_j$ the number of buses in the system. $g_j(x_i(t), \Bar{V}_j(t))$ describes the network and can also be written as
\begin{equation}
    0 = \Bar{Y} \Bar{V} - \Bar{I} = g_j(x_i(t), \Bar{V}_i(t)), \label{eq:ohmslaw}
\end{equation}
where $\Bar{Y}$ is the complex admittance matrix. 

The studied IEEE test systems consist of synchronous generators, condensers and loads distributed among its buses. The synchronous machines are modeled with the state space model and stator circuit described by \eqref{eq:statespace} and \eqref{eq:alggen}, as described in \cite{sauerandpi}.
\begin{equation}\label{eq:statespace}
{ \scriptsize
    \begin{bmatrix}
     T'_{do}  \\ 
     T'_{qo}   \\
     1 \\
     2H
    \end{bmatrix} \frac{d}{dt}
    \begin{bmatrix}
     E'_q \\
     E'_d \\
     \delta \\
     \Delta \omega
    \end{bmatrix} = 
    \begin{bmatrix}
     -E'_q - (X_d-X'_d)I_d +E_{fd}  \\ 
     -E'_d + (X_q-X'_q)I_q \\
     2\pi f \Delta \omega \\
     P_m-E'_dI_d-E'_qI_q-(X'_q-X'_d)I_dI_q - D \Delta \omega
    \end{bmatrix}
}
\end{equation}
\begin{equation} \label{eq:alggen}
{\footnotesize
    \begin{bmatrix}
        I_d \\ I_q
    \end{bmatrix} =
    \begin{bmatrix}
        R_s & -X'_q \\
        X'_d & R_s
    \end{bmatrix}^{-1}
    \begin{bmatrix}
        E'_d-V \sin (\delta - \theta) \\
        E'_q-V \cos (\delta - \theta)
    \end{bmatrix},
    }
\end{equation}
where $\{E'_q, E'_d, \delta, \Delta \omega\}$ are the differential states and $\{I_d, I_q, V, \theta\}$ the algebraic variables. For this study, we simplify the machine model by setting $X'_q=X'_d$ and finding the integral manifold where $E'_q$ and $E'_d$ remain constant. Here, \eqref{eq:statespace} describes the state update function \eqref{daedif}, and is approximated by either a trapezoidal rule as in \eqref{daetrapzdif} or by a PINN approximation as in \eqref{eq:pinnformu}. \eqref{eq:alggen} represents the internal circuit represented by \eqref{daecurrents}. Each dynamic component is added together with the network equations \eqref{eq:ohmslaw}. The formulation is developed using a per unit system.

\Cref{tab:machineparam} summarizes the structure of the studied systems. There are two types of dynamic components (DyC): synchronous generators (SG) and condensers (SYNC) \cite{parameters}. These components are connected to the buses specified for each system. The synchronous generator \textbf{SG*}, present in all studied systems, will be the dynamic component replaced by a PINN. Thus, we train a single PINN that represents its parameters and integrate it into the simulation of all systems under different network disturbances. The following numerical analysis shows the performance gains of integrating PINNs into simulations. The parameters of all machines are available in \cite{github_PINN}.
\begin{table}[!ht]
\centering
\caption{Systems' components, loads and lines description}\label{tab:machineparam} 
\begin{tabular}{ccccccc}
\toprule
\textbf{System} & \textbf{\#DyC} & \textbf{SG} & \textbf{SG*} & \textbf{SYNC} &  \textbf{\#Loads} & \textbf{\#Lines}  \\
\midrule
  9-bus  & 3 & \{1,2\} & \{3\} & -  &  3 & 9  \\
  14-bus & 5 & \{1\}   & \{2\}  & \{3,6,8\} &  11 & 20 \\
  30-bus & 6 & \{1\}  & \{2\} & \{5,8,11,13\} &  22 & 41 \\
  \bottomrule
\end{tabular}
\end{table}

\subsection{PINN implementation}

The power system simulation framework and the PINN training are developed in Python and leverage the NumPy and Torch packages \cite{numpy, torch}. The training is implemented using an NVIDIA Tesla A100 GPU \cite{DTU_DCC_resource}. The loss function is optimized with the Adam optimizer for $10^6$ epochs. The learning rate is reduced using a delayed exponential decay function. The trained PINN consists of 64 neurons for each of the three hidden layers. We use the \textit{tanh} activation function. 

\subsection{Simulation algorithm and disturbances}

Our subsequent analysis compares the simulation performance of two algorithms, the \textit{traditional} and the \textit{hybrid}. Both algorithms are identical. They compute the initial conditions at $t=0$ using a power flow solver and simulate over time using the Newton-Raphson method at each time step. The only difference lies  in the method used to approximate the dynamics of the machine SG*. In the \textit{traditional} algorithm, the trapezoidal rule is used for all machines, including SG*. In contrast, the \textit{hybrid} algorithm employs the trained PINN to model the dynamics of the machine SG*.

The simulation results are obtained with only one single PINN trained for the machine SG*, which is present in every system. This showcases the modularity of the algorithm, where PINNs for different models can be trained and included in the simulation in a plug-and-play manner. Since the formulation is fully compatible, there are no restrictions on how many components or which ones are captured by PINNs. The captured ranges and parameters of the PINN model used for all simulations are summarized in \Cref{tab:ranges_input}.
\begin{table}[!ht]
\centering
\caption{Input domain of initial conditions $\mathcal{X}_x$ and $\mathcal{X}_y$ in p.u.}\label{tab:ranges_input}
\resizebox{\columnwidth}{!}{%
\begin{tabular}{ccccccc}
\toprule
\bm{$\Delta t\, [ms]$} & \bm{$E'_{q,i}$} & \bm{$E'_{d,i}$} & \bm{$\delta_{i}$} & \bm{$\Delta \omega_{i}$} & \bm{$I_{d,i}^{0,\Delta t}$} & \bm{$I_{q,i}^{0,\Delta t}$} \\
\midrule
$[1, 40]$ & $[0.4, 1.2]$& $[0.4, 1.2]$  & $[-\pi, \pi)$ & $[-0.02, 0.02]$& $[0.4, 0.8]$ & $[0.4, 0.8]$ \\
\midrule
\bm{$H$} & \bm{$D$} & \bm{$X_d$} & \bm{$X'_d$} & \bm{$R_s$} & \bm{$T_m$} & \bm{$E_{fd}$} \\
\midrule
$3.01$ & $0.903$& $1.3125$  & $0.1813$ & $0.0$& $0.85$ & $1.04$ \\
\bottomrule
\end{tabular}
}
\end{table}

\subsection{Simulation results}
This section compares the performance of the two simulation algorithms presented, i.e., the \textit{traditional} solver, which only uses trapezoidal rules for integration, and the \textit{hybrid} solver, which incorporates the PINN to model one of the present machines, SG*. First, we analyze the accuracy improvements achieved by incorporating the PINN into the IEEE 30-bus system while simulating three different network disturbances. The disturbances considered are: an increase in load at bus 20 by 0.08 pu ($\text{L}^{20}_{+0.08}$), a reduction in mechanical torque at bus one by 0.3 pu ($\text{G}^1_{-0.3}$), and a reduction in load at bus 12 by 0.11 pu ($\text{L}^{12}_{-0.11}$). All simulations consider the network disturbances at $t_{fault} =0.2$ s and use a time step size of $\Delta t = 20$ ms. The exact trajectory is computed with $\Delta t = 1$ ms. The results are verified by comparing their converged solution with the power flow tool from PowerModels.jl \cite{coffrin}.

\Cref{tab:average} shows the accuracy improvement of all variables using the hybrid solver against the traditional one. Variables labeled as $x^*$ show the performance improvement of the replaced machine SG*, and $x^{\text{avg}}$ show the average performance improvement of all the other components. The largest accuracy improvement appears in the variables associated with the machine modeled by the PINN, as the trained PINN yields more accurate approximations than the trapezoidal rule. However, the overall system also benefits from a more accurate representation of SG* as seen in \Cref{tab:average}. 
\begin{table}[!ht]
\centering
\caption{Average accuracy improvements for the 30-bus system (in \%)}\label{tab:average} 
  \begin{tabular}{cccccccc}
    \toprule
    \multirow{2}{*}{\textbf{Simulation}} &
      \multicolumn{2}{c}{\bm{$\delta$}} &
      \multicolumn{2}{c}{\bm{$\omega$}} &
      \multicolumn{2}{c}{\bm{$I_{d-q}$}} & 
       \\
      & \bm{$\delta^{avg}$} & \bm{$\delta^*$} & \bm{$\omega^{\text{avg}}$} & \bm{$\omega^*$} & \bm{$I_{d-q}^{\text{avg}}$} & \bm{$I_{d-q}^*$} & \bm{$V_m^{\text{avg}}$}\\
      \midrule
    $\text{L}^{20}_{+0.08}$ & 3.7 & 39.2 & 2.2 & 38.9 & 6.0 & 38.5 & 4.5 \\ [0.2cm]
    $\text{G}^1_{-0.3}$ & 6.4 & 39.7 & 4.1 & 39.5 & 8.6 & 38.8 & 3.3 \\[0.2cm]
    $\text{L}^{12}_{-0.11}$ & 5.6 & 34.4 & 3.6 & 33.3 & 8.1 & 33.7 & 4.9 \\
    \bottomrule
  \end{tabular}
\end{table}

\begin{figure}[!ht]
    \centering
    \includegraphics[width=\linewidth]{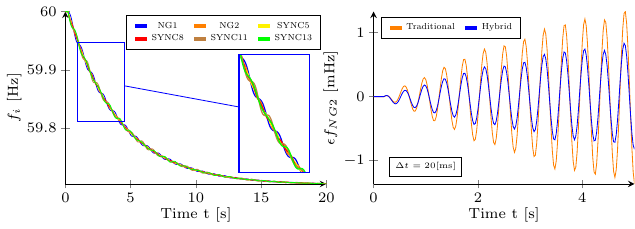}
    \caption{Time evolution of the IEEE 30-bus system after a sudden step in the load of bus 20. On the left, the machine frequencies evolve to convergence. On the right, the first 5-second error evolutions of the frequency $f_{NG2}$, modeled with a trapezoidal rule or PINN inside the system simulation.}
    \label{fig:omegasanalysis}
\end{figure}

\Cref{fig:omegasanalysis} depicts the frequency evolution in the IEEE 30-bus system after a step load increase in bus 20 of 8 MW. The right plot shows an example of how the accumulated errors of the hybrid solver throughout the system variables are smaller and propagate slower than with the traditional solver.

The same PINN model previously applied to the synchronous machine SG* is used across all studied IEEE test systems. In this analysis, we consider a 100\% step increase in the load at the last bus of each system. The results, presented in \Cref{tab:difsys}, demonstrate that PINNs significantly enhance the accuracy of the modeled component while also substantially improving the accuracy of the rest of the system. These findings establish PINNs as a deep learning method capable of enhancing the performance of current simulation solvers.

\begin{table}[!ht]
\centering
\caption{Average accuracy improvements for different systems (in \%)}\label{tab:difsys} 
  \begin{tabular}{cccccccc}
    \toprule
    \multirow{2}{*}{\textbf{System}} &
      \multicolumn{2}{c}{\bm{$\delta$}} &
      \multicolumn{2}{c}{\bm{$\omega$}} &
      \multicolumn{2}{c}{\bm{$I_{d-q}$}} & 
       \\
      & \bm{$\delta^{avg}$} & \bm{$\delta^*$} & \bm{$\omega^{\text{avg}}$} & \bm{$\omega^*$} & \bm{$I_{d-q}^{\text{avg}}$} & \bm{$I_{d-q}^*$} & \bm{$V_m^{\text{avg}}$}\\
      \midrule
    IEEE 9 & 12.8 & 40.3 & 9.5 & 39.9 & 17.9 & 40.3 & 13.1 \\
    IEEE 14 & 2.8 & 44.4 & 0.7 & 40.5 & 8.2 & 40.8 & 2.5 \\
    IEEE 30 & 4.0 & 52.3 & 1.8 & 45.6 & 8.7 & 51.6 & 9.2 \\
    \bottomrule
  \end{tabular}
\end{table}




\section{Conclusions} \label{sec5}
This paper presents a new approach to modularly integrate Physics-Informed Neural Networks (PINNs) into time-domain simulations. 
Initially, we train a PINN to accurately model the dynamics of a single component across a wide range of operating conditions, taking its interaction with the system into account. The trained PINN demonstrates improved accuracy in approximating power system component dynamics, particularly for larger time step sizes.
Subsequently, we incorporate PINNs as an alternative to traditional numerical methods within the dynamic simulation framework. 
We show that the proposed method significantly enhances simulation accuracy by simulating several network disturbances in the IEEE 9-, 14-, and 30-bus systems.
This integration opens the door to a better synergy between time-domain simulations and the emerging machine-learning methods to boost simulating accuracy and speed.
Future work will capture more dynamic components and adapt the formulation to the electromagnetic transients framework.


\bibliographystyle{IEEEtran}
\bibliography{references.bib}

\begin{thebibliography}{10}
\providecommand{\url}[1]{#1}
\csname url@samestyle\endcsname
\providecommand{\newblock}{\relax}
\providecommand{\bibinfo}[2]{#2}
\providecommand{\BIBentrySTDinterwordspacing}{\spaceskip=0pt\relax}
\providecommand{\BIBentryALTinterwordstretchfactor}{4}
\providecommand{\BIBentryALTinterwordspacing}{\spaceskip=\fontdimen2\font plus
\BIBentryALTinterwordstretchfactor\fontdimen3\font minus \fontdimen4\font\relax}
\providecommand{\BIBforeignlanguage}[2]{{%
\expandafter\ifx\csname l@#1\endcsname\relax
\typeout{** WARNING: IEEEtran.bst: No hyphenation pattern has been}%
\typeout{** loaded for the language `#1'. Using the pattern for}%
\typeout{** the default language instead.}%
\else
\language=\csname l@#1\endcsname
\fi
#2}}
\providecommand{\BIBdecl}{\relax}
\BIBdecl

\bibitem{kundur}
P.~Kundur, \emph{Power System Stability and Control}.\hskip 1em plus 0.5em minus 0.4em\relax McGraw-Hill, 1994.

\bibitem{sauerandpi}
P.~Sauer and M.~Pai, \emph{\BIBforeignlanguage{English (US)}{Power System Dynamics and Stability}}, 1st~ed.\hskip 1em plus 0.5em minus 0.4em\relax Prentice Hall, 1998.

\bibitem{stott}
B.~Stott, ``Power system dynamic response calculations,'' \emph{Proceedings of the IEEE}, vol.~67, no.~2, pp. 219--241, 1979.

\bibitem{aristidou}
P.~Aristidou \emph{et~al.}, ``Power system dynamic simulations using a parallel two-level schur-complement decomposition,'' \emph{IEEE Transactions on Power Systems}, vol.~31, no.~5, pp. 3984--3995, 2016.

\bibitem{modelreduction}
D.~Chaniotis and M.~Pai, ``Model reduction in power systems using krylov subspace methods,'' in \emph{IEEE Power Engineering Society General Meeting, 2005}, 2005, pp. 1412 Vol. 2--.

\bibitem{universal}
M.~Leshno, V.~Y. Lin, A.~Pinkus, and S.~Schocken, ``Multilayer feedforward networks with a nonpolynomial activation function can approximate any function,'' \emph{Neural Networks}, vol.~6, pp. 861--867, 1993.

\bibitem{raissi}
M.~Raissi, P.~Perdikaris, and G.~Karniadakis, ``Physics-informed neural networks: A deep learning framework for solving forward and inverse problems involving nonlinear partial differential equations,'' \emph{Journal of Computational Physics}, vol. 378, pp. 686--707, 2019.

\bibitem{resnet}
K.~He, X.~Zhang, S.~Ren, and J.~Sun, ``Deep residual learning for image recognition,'' in \emph{2016 IEEE Conference on Computer Vision and Pattern Recognition (CVPR)}, 2016, pp. 770--778.

\bibitem{piml}
G.~Karniadakis, I.~Kevrekidis, L.~Lu \emph{et~al.}, ``Physics-informed machine learning,'' \emph{Nat Rev Phys 3}, pp. 422--440, 2021.

\bibitem{ann}
I.~Lagaris, A.~Likas, and D.~Fotiadis, ``Artificial neural networks for solving ordinary and partial differential equations,'' \emph{IEEE Transactions on Neural Networks}, vol.~9, no.~5, pp. 987--1000, 1998.

\bibitem{misyris}
G.~S. Misyris, A.~Venzke, and S.~Chatzivasileiadis, ``Physics-informed neural networks for power systems,'' in \emph{2020 IEEE Power \& Energy Society General Meeting (PESGM)}, 2020, pp. 1--5.

\bibitem{apppinns}
B.~Huang and J.~Wang, ``Applications of physics-informed neural networks in power systems - a review,'' \emph{IEEE Transactions on Power Systems}, vol.~38, no.~1, pp. 572--588, 2023.

\bibitem{jochenepsr}
J.~Stiasny and S.~Chatzivasileiadis, ``Physics-informed neural networks for time-domain simulations: Accuracy, computational cost, and flexibility,'' \emph{Electric Power Systems Research}, vol. 224, p. 109748, 2023.

\bibitem{daepinn}
C.~Moya and G.~Lin, ``Dae-pinn: a physics-informed neural network model for simulating differential algebraic equations with application to power networks,'' \emph{Neural Comput \& Applic 35}, p. 3789–3804, 2023.

\bibitem{kyri}
M.~Mohammadian, K.~Baker, and F.~Fioretto, ``Gradient-enhanced physics-informed neural networks for power systems operational support,'' \emph{Electric Power Systems Research}, vol. 223, p. 109551, 2023.

\bibitem{pinnsim}
J.~Stiasny, B.~Zhang, and S.~Chatzivasileiadis, ``Pinnsim: A simulator for power system dynamics based on physics-informed neural networks,'' \emph{Electric Power Systems Research}, vol. 235, p. 110796, 2024.

\bibitem{plugandplay}
\BIBentryALTinterwordspacing
I.~V. Nadal, J.~Stiasny, and S.~Chatzivasileiadis, ``Physics-informed neural networks: a plug and play integration into power system dynamic simulations,'' 2025. [Online]. Available: \url{https://arxiv.org/abs/2404.13325}
\BIBentrySTDinterwordspacing

\bibitem{github_PINN}
I.~Ventura, R.~Nellikkath, and S.~Chatzivasileiadis, ``Pinns for ieee systems,'' \url{https://github.com/ignvenad/PINNs-for-IEEE-systems}, 2025.

\bibitem{index1}
M.~La~Scala and A.~Bose, ``Relaxation/newton methods for concurrent time step solution of differential-algebraic equations in power system dynamic simulations,'' \emph{IEEE Transactions on Circuits and Systems I: Fundamental Theory and Applications}, vol.~40, pp. 317--330, 1993.

\bibitem{iserles}
A.~Iserles, \emph{A first course in the numerical analysis of differential equations}.\hskip 1em plus 0.5em minus 0.4em\relax USA: Cambridge University Press, 1996.

\bibitem{Goodfellow}
I.~Goodfellow, Y.~Bengio, and A.~Courville, \emph{Deep Learning}.\hskip 1em plus 0.5em minus 0.4em\relax MIT Press, 2016, \url{http://www.deeplearningbook.org}.

\bibitem{illinois}
I.~C. for~a Smarter Electric~Grid, 2013, [Online]. Available: \url{http://publish.illinois.edu/smartergrid/}.

\bibitem{parameters}
P.~Demetriou \emph{et~al.}, ``Dynamic ieee test systems for transient analysis,'' \emph{IEEE Systems Journal}, vol.~11, no.~4, 2017.

\bibitem{numpy}
C.~R. Harris \emph{et~al.}, ``Array programming with {NumPy},'' \emph{Nature}, vol. 585, no. 7825, pp. 357--362, Sep. 2020.

\bibitem{torch}
A.~Paszke \emph{et~al.}, ``Pytorch: An imperative style, high-performance deep learning library,'' in \emph{Advances in Neural Information Processing Systems}, vol.~32.\hskip 1em plus 0.5em minus 0.4em\relax Curran Associates, Inc., 2019.

\bibitem{DTU_DCC_resource}
\BIBentryALTinterwordspacing
{DTU Computing Center}, ``{DTU Computing Center resources},'' 2024. [Online]. Available: \url{https://doi.org/10.48714/DTU.HPC.0001}
\BIBentrySTDinterwordspacing

\bibitem{coffrin}
C.~Coffrin \emph{et~al.}, ``Powermodels.jl: An open-source framework for exploring power flow formulations,'' in \emph{2018 Power Systems Computation Conference (PSCC)}, June 2018, pp. 1--8.

\end{thebibliography}

\end{document}